\documentclass[aps,twocolumn,raggedbottom,prl,showpacs,nobalancelastpage,amssymb,groupedaddress]{revtex4}
\usepackage{epsfig}
\usepackage{verbatim}	 
\usepackage{amssymb}
\usepackage{bbold}
\usepackage{amsmath}
\usepackage{psfrag}
\usepackage{color}
\usepackage{ulem}
\makeatletter
\renewcommand*\env@matrix[1][\arraystretch]{%
  \edef\arraystretch{#1}%
  \hskip -\arraycolsep
  \let\@ifnextchar\new@ifnextchar
  \array{*\c@MaxMatrixCols c}}
\makeatother

\newcommand{\bi}{\bibitem}

\newcommand{\Lc}{\mathcal{L}}
\newcommand{\Tc}{\mathcal{T}}
\newcommand{\Ac}{\mathcal{A}}

\newcommand{\eye}{\mathbb{1}}
\newcommand{\Tpn}{T_{\rm p}^{\rm (noise)}}
\newcommand{\Tr}{{\rm Tr}}

\begin{document}

\hsize\textwidth\columnwidth\hsize\csname@twocolumnfalse\endcsname

\title{Local Temperature of Out-of-Equilibrium Quantum Electron
Systems}

\author{J. Meair$^1$, J.~P. Bergfield$^2$, C.~A. Stafford$^1$, and Ph.~Jacquod$^{1,3}$}

\affiliation{$^1$Department of Physics, University of Arizona, 1118 E.\ 4th St., Tucson, AZ, 85721 \\
$^2$ Department of Chemistry, Northwestern University, 2145 Sheridan Road, Evanston, IL, 60208 \\
$^3$ College of Optical Sciences, University of Arizona, Tucson, AZ 85721}

\vskip1.5truecm
\begin{abstract}
We show how the local temperature of out-of-equilibrium,  quantum electron
systems can be consistently defined with the help of an external voltage and temperature probe. 
We determine sufficient conditions under which the temperature measured by the probe 
(i) is independent of details of the system-probe coupling, (ii) is equal to the 
temperature obtained from an independent current-noise measurement, (iii) satisfies the transitivity 
condition expressed by the zeroth law of 
thermodynamics, and  (iv) is consistent with Carnot's 
theorem. This local temperature therefore characterizes the 
system in the common sense of equilibrium thermodynamics, but remains well defined 
even in out-of-equilibrium situations with no local equilibrium.
\end{abstract}

\maketitle
{\bf Introduction.} Thermodynamics characterizes systems at equilibrium via equations of state that depend on a
few macroscopic variables, in particular the temperature. The latter, not being an observable in the dynamical sense, can be defined
in various ways. The zeroth law differentiates between
classes of thermodynamic states with different temperatures, and an absolute temperature scale is introduced by the
second law via Carnot's theorem. 
Maxwell relations express the temperature
as derivatives of thermodynamic potentials with respect to the entropy. 
Fluctuation-dissipation theorems finally 
relate the temperature to equilibrium fluctuations of observables via associated response coefficients~\cite{Kardar}. 
In equilibrium, these definitions are consistent with one another. 

The framework of thermodynamics, and the concept of temperature in particular have 
been 
extended to nonequilibrium 
systems under the assumption of local equilibrium~\cite{Groo62}. However, it has proven far more challenging to  
generalize the temperature concept to 
systems where the local equilibrium hypothesis does not 
hold~\cite{Cug11,Cas03}. Without
local equilibrium,  different temperatures are commonly obtained by different
measurement protocols~\cite{Cas03}. The consensus is accordingly that trying to extend the 
concept of temperature to out-of-equilibrium thermodynamics can at best deliver an operational definition.

In this manuscript, we revisit and shed new light on this fundamental issue. We focus our investigations on 
coupled electric and thermal transport in quantum conductors
brought out of equilibrium by voltage and temperature biases. 
We show that, under certain conditions which we specify,
a local temperature can be consistently defined in this out-of-equilibrium system
in the sense that: (i) the temperature is insensitive to details of the measurement protocol; (ii) the same temperature is given by at least two completely different measurements (in our case a direct thermal measurement and an
electric noise measurement); (iii) two systems independently at equilibrium with 
a third one are also at equilibrium with one another; 
and (iv) the measured temperature is absolute in the sense of Carnot's theorem. 

Our approach is inspired by the experimental thermometry technique of 
scanning thermal microscopy~\cite{Maj99}, whose resolution has recently been brought down to 
the nanometer range~\cite{Kim12}.  
The system's local temperature is defined via an external local probe  weakly coupled 
to the system via a tunnel barrier~\cite{Ber12}.
At its other end, the probe is connected to a macroscopic
reservoir whose chemical potential and temperature are set such that neither electric nor heat current flow
between the probe and the system. 
The probe is thus in local equilibrium with a system that is itself
not at equilibrium. In linear response, the probe temperature guaranteeing this local equilibrium is unique,
and we show that this temperature locally characterizes the system in the sense of points (i)--(iv). The 
local temperature remains consistently defined even when there is no local equilibrium in the system itself.
In particular,
quantum interference effects that destroy equilibrium on scales comparable to the Fermi wavelength do not 
alter the consistency of our definition.

The physics of electronic transport in quantum coherent systems coupled
to external probes dates back to B\"uttiker's work on dephasing~\cite{But88}. Probes have been used to 
calculate local electronic distributions~\cite{Gra97} and local spin accumulations~\cite{Jac10} in such systems. 
The approach has recently been extended to probe thermometry in voltage- and/or temperature-biased
structures~\cite{Jac09,Dub09,Dubi09b,Dubi09a,Cas10,Ber12},
with several investigations focusing on Fourier's law~\cite{Jac09,Dub09,Dubi09b,Dubi09a,Ber12}.  
Ref.~\cite{Cas10} investigated the probe temperature of AC driven 
systems in the weak-driving, low-frequency limit. Neglecting thermoelectric effects, it was found that the temperature measured by the probe is the same as the one extracted from a local 
fluctuation-dissipation relation.  Recently, local temperature measurements were investigated in the more general case including 
thermoelectric effects, where a closed-form analytic expression for the temperature was found for open electrical circuits~\cite{Ber12}.

Refs.~\cite{Maj99,Kim12,Jac09,Dub09,Dubi09b,Dubi09a,Cas10,Ber12}
considered the probe temperature as an operational definition
of the local temperature of the sample, without examining 
whether this definition satisfies conditions obeyed by a temperature in the thermodynamic sense.
Here, we fill this gap by
investigating the  fundamental issue of whether a local temperature can 
be consistently defined in quantum electron systems out of equilibrium and under what
conditions this temperature is the same as that measured by an external probe. 

\begin{figure}[ht]
\psfrag{m1}{$\mu_1$}
\psfrag{m2}{$\mu_2$}
\psfrag{mp}{$\mu_{\rm p}$}
\psfrag{T1}{$T_1$}
\psfrag{T2}{$T_2$}
\psfrag{Tp}{$T_{\rm p}$}
\includegraphics[width=3.2in]{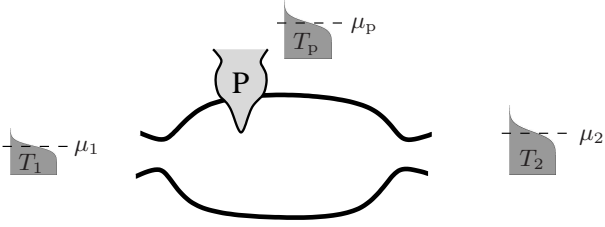}
\caption{Schematic of a mesoscopic conductor connected to two electron reservoirs via transport leads, and 
to a third reservoir via a weakly coupled probe. The chemical potentials and temperatures in all three
reservoirs are indicated. 
}\label{fig:probe}
\end{figure}

{\bf Model and scattering approach to transport.} 
The system we investigate is sketched in Fig.~\ref{fig:probe}. It is an 
electronic system connected to 
external reservoirs where electrons are thermalized and have a well defined Fermi-Dirac distribution.
One of the reservoirs is
coupled to the conductor via a tunnel probe, and both its temperature $T_{\rm p}$ and chemical potential 
$\mu_{\rm p}$ are set such that neither electric nor heat current flow between the conductor and the
probe. Thus the conductor and the probe are in local equilibrium, even though the conductor itself is not
and may be traversed by heat and electric currents.
We consider transport mediated solely by electrons.

We use the scattering approach to transport~\cite{But88,Imry86,But90} which, in linear response, 
expresses the electrical current, $I = e I^{(0)}$, and heat current, $J=I^{(1)}$, flowing from reservoir $\alpha$ into the conductor as
\begin{eqnarray}
\label{eq:linear}
I_\alpha^{(\nu)}
= 
\sum_\beta \left[
\Lc_{\alpha \beta}^{(\nu,1)} (\mu_\beta-\mu_0) 
+
\Lc_{\alpha \beta}^{(\nu+1,1)} \left(\frac{T_\beta-T_0}{T_0}\right)\right] \, .
\end{eqnarray}
Here, $e$ is the electron charge and $\mu_0$ and $T_0$ are the base chemical potential and temperature, which
we fix at their equilibrium values. The sum over $\beta$ runs over the transport as well as the probe terminals.
The linear response coefficients $\Lc$
are given by~\cite{But90}
\begin{align}\label{eq:Ldef}
\Lc_{\alpha \beta}^{(\nu,\lambda)} 
&= 
\frac{1}{h}\int  {\rm d} E \;(E-\mu_0)^\nu 
\;(-1)^\lambda \;(\partial_E^\lambda f) \;A_{\alpha \beta}(E)\;\;,\end{align}
where $E$ is the energy, $f=[1+{\rm exp}\{(E-\mu_0)/k_B T_0\}]^{-1}$ 
is the equilibrium Fermi-Dirac distribution 
and $A_{\alpha \beta}(E) = 2 N_\alpha(E) 
\delta_{\alpha \beta} - \Tc_{\alpha \beta}(E)$, where $N_\alpha$ is the number 
of transport channels in lead $\alpha$ 
and $\Tc_{\alpha \beta}=\Tr\left[s^\dagger_{\alpha \beta} s_{\alpha \beta}\right]$ is
given by the sub-block  $s_{\alpha \beta}$ of the scattering matrix connecting lead $\beta$ to $\alpha$. The trace in this latter expression
is taken over both the electron spin and  the transport channels.  For the currents, only coefficients with $\lambda  = 1$ 
in Eq. (2) matter; 
however, those with $\lambda = 2$ appear in expressions for the current noise.
Charge conservation and gauge invariance are 
expressed mathematically by the unitarity of the scattering matrix, 
$\sum_{\alpha} \Lc_{\alpha \beta}^{(\nu, \lambda)}=0=\sum_{\beta} \Lc_{\alpha \beta}^{(\nu, \lambda)}$. 

\begin{figure}[h]
\includegraphics[width=3.2in]{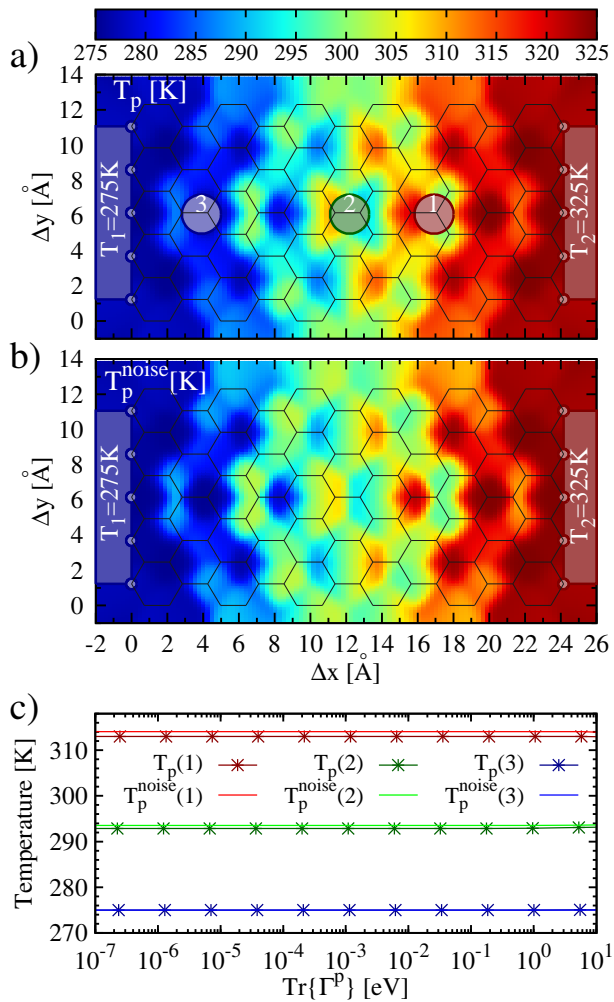}
\caption{(Color online) 
Local temperature of an armchair graphene 
nanoribbon probed by an atomically-sharp Pt tip scanned 3.5\AA\ above the graphene plane.
Here $\mu_0
= -0.15\mbox{eV}$ from the Dirac point, and a
thermal bias $\Delta T=50\mbox{K}$ is 
applied between hot and cold electrodes forming an open electrical circuit.
(a) $T_{\rm p}$ calculated from Eq.\ (\ref{eq:Tpgen}); (b) $\Tpn=T_{\rm p}(1+\chi)$ calculated from Eq.\ (\ref{eq:chi}); (c) $T_{\rm p}$ and $\Tpn$ at the
three points indicated in panel (a)
as functions of the tip-sample coupling $\mbox{Tr}\{\Gamma^p\}$,
which we vary with an artificial scaling factor multiplying the tunneling-width matrix $\Gamma^p$.
At this scan height, the tip-sample coupling is mediated by two dominant transmission channels, with an intrinsic
$\mbox{Tr}\{\Gamma^p\} \in [3.6\mu\mbox{eV},24\mbox{meV}]$ over the image.
}
\label{fig:graphene}
\end{figure}

{\bf Probe temperature.} 
In linear response, there is a single chemical potential $\mu_{\rm p}$ and temperature $T_{\rm p}$ ensuring
$I_{\rm p}^{(0,1)}=0$. 
The probe temperature is 
\begin{align}\label{eq:Tpgen}
\frac{T_{\rm p}}{T_0}&=\kappa_{\rm pp}^{-1}\sum_{\alpha \neq {\rm p}} 
\left[ 
\left(\Lc_{pp}^{(0,1)}\Lc_{p\alpha}^{(1,1)}-\Lc_{pp}^{(1,1)}\Lc_{p\alpha}^{(0,1)}\right)\mu_\alpha \right. \nonumber \\
&\qquad\qquad\;\left.+
\left(\Lc_{pp}^{(0,1)}\Lc_{p\alpha}^{(2,1)}-\Lc_{pp}^{(1,1)}\Lc_{p\alpha}^{(1,1)}\right)\frac{T_\alpha}{T_0} 
\right]\; ,
\end{align}
with
$\kappa_{pp}=\left(\Lc_{pp}^{(1,1)}\right)^2-\Lc_{pp}^{(0,1)}\Lc_{pp}^{(2,1)}$.
Eq.~(\ref{eq:Tpgen}) applies to general thermoelectric 
circuits (with voltage biases, temperature biases, or both)
and agrees with the previous result~\cite{Ber12} for the specific case of heat transport in an open electrical circuit.
An expression similar to Eq.\ (\ref{eq:Tpgen}) was derived in Ref.~\cite{San11}.
As an example, a plot of $T_{\rm p}$ for an armchair graphene nanoribbon with a thermal bias of
$\Delta T=50\mbox{K}$ is shown in Fig.~\ref{fig:graphene}(a).

{\bf Dependence of $T_{\rm p}$ on probe-system coupling.}
Let us first consider the case of
single-channel probe-system coupling.
For this case, it was shown in Ref.~\cite{Jac10} that
the sub-blocks $s_{\alpha p}$ (and $s_{p \alpha}$)  of the scattering matrix factorize as
$s_{\alpha p} = \gamma(E) \tilde{s}_{\alpha p}$, with the (possibly energy-dependent)
coupling between system and probe encoded in $\gamma(E)$ only. Thus one has
$\Tc_{\alpha p}(E) = |\gamma(E)|^2 \tilde{\Tc}_{\alpha p}(E)$. Next, we rewrite
Eq.~(\ref{eq:Ldef}), expanding the coefficients
$A_{\alpha \beta}$ about the equilibrium chemical potential as
\begin{eqnarray}\label{eq:expand}
\Lc_{\alpha \beta}^{(\nu,1)} 
&=& 
\frac{1}{h}\int  {\rm d}E \;(E-\mu_0)^\nu  \;(-\partial_E f) \times   \; \left[A_{\alpha \beta}(\mu_0)  \right. \nonumber \\
&& \left.  + (E-\mu_0) \, [\partial_E A_{\alpha \beta}(\mu_0)]
+ {\mathcal O}\{(E-\mu_0)^2\} \right] \,  . \qquad
\end{eqnarray}
This expansion is consistent with the Sommerfeld expansion leading, e.g.\
to Mott's relation for the thermopower. 
By symmetry, we have that 
the $A_{\alpha \beta}(\mu_0)$-term contributes when $\nu=0$ while 
the $(E-\mu_0) \, [\partial_E A_{\alpha \beta}(\mu_0)]$-term
contributes when $\nu=1$. 
Setting $\beta=p$, we factorize the coefficients
$A$ in the integral in Eq.~(\ref{eq:expand}) and write 
$\partial_E A_{\alpha p}(\mu_0)=|\gamma(\mu_0)|^2 \, \partial_E\tilde{A}_{\alpha p}(\mu_0) +  \tilde{A}_{\alpha p}(\mu_0) \,
 \partial_E |\gamma(\mu_0)|^2$. When
 $ \tilde{A}_{\alpha p}(\mu_0) \,
 \partial_E |\gamma(\mu_0)|^2 \ll |\gamma(\mu_0)|^2 [\partial_E\tilde{A}_{\alpha p}(\mu_0)]$,
 $|\gamma(\mu_0)|^2$ factors out of 
 both the numerator and denominator of Eq.~(\ref{eq:Tpgen}), in which case the temperature measured by the
 probe is independent of the strength and energy-dependence of the coupling between system and probe. 
 Thus, as long as linear thermoelectric effects involving transmission from and to 
 the probe are dominated by the energy-dependence of 
 transmission coefficients inside the system [as opposed to the energy-dependence of $\gamma(E)$]
 and when the system-probe coupling proceeds via a single transport channel, $T_{\rm p}$ is independent
 of $\gamma(E)$. This condition is typically 
satisfied for tunneling probes, which 
have transmissions that
vary over an 
energy scale in the $\Delta \sim eV$ range. Their energy-dependence can therefore
safely be ignored when probing nanoelectronic systems whose transmission fluctuates over a scale set by
the Thouless energy $E_{\rm Th} \lesssim 10^{-1} eV$ for 
a typical system of linear size $L \gtrsim 10$ nm. 
  
{\bf Current noise temperature.} The condition 
$I_{\rm p}^{(0,1)}=0$ leading to 
Eq.~(\ref{eq:Tpgen}) means that time-averaged currents 
into the probe
vanish; however, they have non-zero 
time-dependent fluctuations. At equilibrium,  
a well-known fluctuation-dissipation relation relates the zero-frequency
electric current noise power to the system's equilibrium temperature, $S = 4 G k_{\rm B}T_0$~\cite{JN}, with the 
linear conductance $G$ of the system.  We next use a  Sommerfeld
expansion to show that a similar fluctuation-dissipation relation exists between
the probe temperature $T_{\rm p}$ and the noise of the electric current between the system and the probe.

In the scattering approach, the electric current cross correlation between terminals 
$\alpha$ and $\beta$ is given by~\cite{Bla00}
\begin{eqnarray}
\frac{S_{\alpha \beta}}{G_0}&=&
\int  dE \; \sum_{\gamma \delta} \;\Tr\left[ \Ac_{\gamma \delta}(\alpha) \Ac_{\delta \gamma}(\beta)\right] \nonumber\\
&&\qquad\times\left[f_\gamma(1-f_\delta)+f_\delta(1-f_\gamma)\right]\;,
\end{eqnarray}
where $G_0=e^2/h$ is the conductance quantum, and 
$\Ac_{\gamma \delta}(\alpha,E) = \eye_\alpha \delta_{\alpha \gamma}\delta_{\alpha \delta} - s_{\alpha \gamma}^\dagger(E) s_{\alpha \delta}(E)$ with the $2N_\alpha \times 2N_\alpha$ identity matrix $\eye_\alpha$.
Within linear response and with Eq.~(\ref{eq:Ldef}), we obtain the current noise in the probe as 
\begin{align}\label{JN}
S_{pp} 
&=
4 \;G(\mu_{\rm p},T_{\rm p}) 
\; k_B\Tpn\;\;,\end{align}
where 
\begin{align}
G(\mu_{\rm p},T_{\rm p})=G_0 \left[\Lc_{pp}^{(0,1)}+\Lc_{pp}^{(0,2)}\vphantom{\left(\frac{T_{\rm p}-T_0}{T_0}\right)}\right.&\left(\mu_{\rm p}-\mu_0\right)\nonumber\\
+\left(\Lc_{pp}^{(1,2)}-\Lc_{pp}^{(0,1)}\right)&\left.\left(\frac{T_{\rm p}-T_0}{T_0}\right)\right]
\end{align}
is the sample-to-probe conductance evaluated at the local electrochemical potential and temperature, and $\Tpn=T_{\rm p} \left(1+\chi\right)$ with
\begin{align}
\chi 
&=
-\frac{1}{2\Lc_{pp}^{(0,1)}}
\sum_{\alpha} 
\left[ 
\Lc_{p \alpha}^{(0,2)} (\mu_\alpha-\mu_0) + \Lc_{p \alpha}^{(1,2)} \left(\frac{T_\alpha-T_0}{T_0}\right) \right]\;\;.
\label{eq:chi}
\end{align}
Eq.~(\ref{JN}) is the Johnson-Nyquist noise for an equilibrium system 
with, however, $\Tpn$ instead of $T_{\rm p}$. 
Clearly, when the system is at equilibrium, $\Tpn=T_{\rm p}$, and a direct calculation using the Sommerfeld expansion
shows  that when the system is biased out of equilibrium, 
$\chi \propto I_{\rm p}^{(1)}$, implying $\Tpn=T_{\rm p}$ when $I_{\rm p}^{(1)}=0$. 
We conclude that the temperature $T_{\rm p}$ measured at the probe is equal to the purely electrically measured
temperature $\Tpn$ in the regime of validity of the Sommerfeld expansion.

In order to illustrate these findings, and to test their validity under somewhat more general conditions typical of a 
realistic tunneling probe, 
we calculate both $T_{\rm p}$ and $\Tpn$ 
for an armchair graphene nanoribbon probed by an atomically-sharp Pt tip (see Fig.~\ref{fig:graphene}). 
The tunnel coupling is mediated by the $s$, $p$, and $d$ orbitals of the apex atom of the tip,
leading to a coupling matrix $\Gamma^p$ between system's modes and probe orbitals. 
We found that the overlap between the Pt orbitals and the C $\pi$-orbitals in graphene
yields two dominant transmission channels into the tip. The connection between the eigenvalues 
of $\Gamma^p$ and the tunnel probabilities $|\gamma_n(E)|^2$ is discussed e.g. in Ref.~\cite{Bee97}.
For details of the model 
used for a scanning thermoelectric probe of graphene, 
see Refs.~\cite{Ber12,Ber13}.

It is apparent from Figs.\ \ref{fig:graphene}(a) and \ref{fig:graphene}(b) that the local temperatures inferred from a direct thermal measurement
and from an independent current noise measurement are almost identical.
Indeed, the maximum value of the discrepancy $\chi$ [cf.\ Eq.\ (\ref{eq:chi})] is just 0.0162.
This agreement is remarkable, especially since there is no local equilibrium, as indicated by 
the short-wavelength coherent spatial oscillations of the temperature. 
Fig.\ \ref{fig:graphene}(c) shows $T_{\rm p}$ and $\Tpn$ at three points within the sample [indicated in Fig.\ \ref{fig:graphene}(a)] 
as functions of the tip-sample coupling.
Despite the fact that the tip-sample coupling is effectively mediated by two transmission channels in this case, 
both $T_{\rm p}$ and $\Tpn$ are seen to be
essentially independent of $\mbox{Tr}\{\Gamma^p\}$ over several orders of magnitude, confirming the analytical argument given above.

\begin{figure}[ht]
\psfrag{m1t1}[][][1.5]{$\substack{\mu_1 \vphantom{\mu_1'}\\\vphantom{T_1'} T_1}$}
\psfrag{m2t2}[][][1.5]{$\substack{\mu_2 \vphantom{\mu_1'}\\\vphantom{T_1'} T_2}$}
\psfrag{m1t1'}[][][1.5]{$\substack{\mu_1'\\ T_1'}$}
\psfrag{m2t2'}[][][1.5]{$\substack{\mu_2'\\ T_2'}$}
\psfrag{mptp}{$\mu_{\rm p}, T_{\rm p}$}
\psfrag{mptp'}{$\mu_{\rm p}', T_{\rm p}'$}
\psfrag{IJ-}{$I_+, J_+$}
\psfrag{IJ+}{$I_-, J_-$}
\psfrag{x1}{${\bf x}_1$}
\psfrag{x2}{${\bf x}_2$}
\includegraphics[width=3.2in]{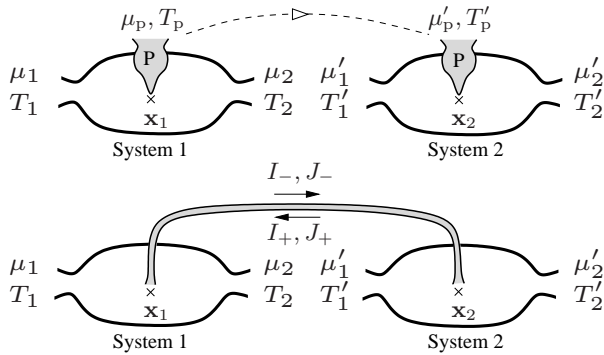}
\caption{Top panel: two systems sequentially probed by the same probe at local positions
${\bf x}_1$ and ${\bf x}_2$. Bottom panel: the same systems as in the top panel
now connected by a 
transmission line locally coupled to the systems at ${\bf x}_1$ and ${\bf x}_2$.}\label{fig:0law}
\end{figure}

{\bf Consistency with the zero$^{\rm \bf th}$ law.} We next use the same probe to sequentially measure the 
local temperature of two different, independent two-terminal systems. We assume that in the absence of bias, the two 
systems are at equilibrium with one another, with in particular the same equilibrium Fermi function, and
that once they are biased, 
there is at least a pair of
positions ${\bf x}_1$ on system 1 and ${\bf x}_2$ on system 2 where the probe measures the same 
$\mu_{\rm p}$ and $T_{\rm p}$. We then 
connect these two points by a transmission line 
with the same transparency $|\gamma(E)|^2$ as the probe (see Fig.~\ref{fig:0law}). 
We assume a weak coupling $\gamma(E)$ 
and treat it to lowest order in perturbation theory,
so that the transmission coefficients between the two systems factorize as
(we use prime indices for system 2) $A_{\alpha \beta'} = -\tilde A_{\alpha p} |\gamma(E)|^2 \tilde A_{p \beta'}$, in terms of the 
transmission coefficients $\tilde{A}$ of the disconnected systems. 

The currents flowing through the neck ($I_+^{(\nu)}$ from system 2 to 
system 1 and $I_-^{(\nu)}$ from 1 to 2) are
\begin{subequations}
\begin{eqnarray}
I_+^{(\nu)} & = & -\frac{1}{h}\int {\rm d} E \;(E-\mu_0)^\nu \sum_{\alpha=1,2} \sum_{\beta'=1',2'} A_{\alpha \beta'}  f_{\beta'} \, , \\
I_-^{(\nu)} & = & -\frac{1}{h}\int {\rm d} E \;(E-\mu_0)^\nu \sum_{\alpha'=1',2'} \sum_{\beta=1,2} A_{\alpha'\beta} f_{\beta} \, . \qquad
\end{eqnarray}
\end{subequations}
Factorizing the transmission coefficients as indicated above, using the unitarity condition
$\sum_\alpha A_{\alpha\beta} = \sum_\beta A_{\alpha\beta} = 0$, the condition of vanishing currents at the probe 
$I_{\rm p}^{(0,1)}=0$, and a Sommerfeld expansion, we obtain the net currents $I^{(\nu)}\equiv I_+^{(\nu)} - I_-^{(\nu)}$
\begin{align}
\begin{bmatrix}[1.5]
 I^{(0)} \\ I^{(1)} 
\end{bmatrix} =
\begin{bmatrix}[1.5]
 \mathcal{M} &
a \, \partial_E \, \mathcal{M}  \\
a \, \partial_E \,  \mathcal{M}  &
a \,  \mathcal{M}
\end{bmatrix} \begin{bmatrix}[1.5]
\mu_{\rm p}'-\mu_{\rm p} \\ \frac{\textstyle T_{\rm p}'-T_{\rm p}}{\textstyle T_0}
\end{bmatrix} \, ,
\end{align}
where $\mathcal{M} = h^{-1} |\gamma|^2 \tilde A_{pp}(\mu_0) \tilde A_{pp}'(\mu_0)$ and $a=(\pi k_B T_0)^2/3$. 
We see that
$I^{(\nu)} = 0$, i.e.\ the two systems, once biased, are at local equilibrium
with each other when connected via points where their probe temperature and chemical 
potential are the same. This brings further consistency to the temperature defined by the probe, in the
sense of the zeroth law of thermodynamics. 

{\bf Consistency with Carnot's theorem.} 
The junction between the system and probe can act as a heat engine
when the temperature and chemical potential 
of the latter are 
biased away from their local equilibrium values $\mu_{\rm p} \rightarrow \mu_{\rm p} + \delta \mu$,
$T_{\rm p} \rightarrow T_{\rm p} + \delta T$. The resulting flow of heat is accompanied by electrical work, and the efficiency 
of the engine is  
$\eta = -I_{\rm p}^{(0)} \delta \mu/ I_{\rm p}^{(1)}$. To linear order, the currents are given by
\begin{equation}
I_{\rm p}^{(\nu)}  =  \Lc_{pp}^{(\nu,1)} \delta \mu + \Lc_{pp}^{(\nu+1,1)} \delta T / T_0 \, .
\end{equation}
They are 
identical
to those for a two-terminal engine in linear response, for which the maximal efficiency  is 
\begin{eqnarray}
\eta_{\rm max} &=& \left(\frac{\sqrt{1+\mathcal{ZT}}-1}{\sqrt{1+\mathcal{ZT}}+1}\right) \frac{
|\delta T|
}{T_0} \, , 
\end{eqnarray}
with a dimensionless figure of
merit
$\mathcal{ZT}^{-1}=\Lc_{pp}^{(0,1)} \Lc_{pp}^{(2,1)}/\left(\Lc_{pp}^{(1,1)}\right)^2-1$. 
We see that, aside from linear transport coefficients,   $\eta_{\rm max}$ depends on the ratio of the 
temperatures 
of the system and probe
only, and therefore defines an absolute temperature scale
in the sense of Carnot's theorem. 

{\bf Conclusion.}  
We have shown that defining the local temperature of a system out of equilibrium via an external thermal probe 
is consistent with both the zeroth and second laws of thermodynamics, as well as 
a fluctuation-dissipation theorem.
Moreover, the temperature is independent of the probe-sample coupling over a wide range of conditions within the tunneling
regime.  Importantly, our findings hold even when 
the system is far from any local equilibrium 
due for instance to quantum interference effects as illustrated in Fig.~\ref{fig:graphene}.

JPB was supported by the Non-Equilibrium Energy Research Center (NERC), an Energy Frontier Research Center funded by DOE-BES under Award DE-SC0000989. CAS was supported by DOE-BES DE-SC0006699.


\begin{thebibliography}{99}

\bi{Kardar} M. Kardar, {\it Statistical Physics of Particles}, Cambridge University Press (New York, 2007).

\bi{Groo62} S.R. de Groot and P. Mazur, {\it Non-Equilibrium Thermodynamics}, North-Holland (Amsterdam, 1962).

\bi{Cug11} L.F. Cugliandolo, J. Phys. A: Math. Theor. {\bf 44}, 483001 (2011).

\bi{Cas03} J. Casas-V\'asquez and D. Jou, Rep. Prog. Phys. {\bf 66}, 1937 (2003).

\bi{Maj99} A. Majumdar, Annu. Rev. Mater. Sci. {\bf 29}, 505 (1999).

\bi{Kim12} K. Kim, W. Jeong, W. Lee, and P. Reddy, ACS Nano {\bf 6}, 4248 (2012).

\bi{Ber12} J.P. Bergfield, S.M. Story, R.C. Stafford, and C.A. Stafford, ACS Nano {\bf 7}, 4429 (2013). 

\bi{Jac09} P.A. Jacquet, J. Stat. Phys. {\bf 134}, 709 (2009); P.A. Jacquet and C.-A. Pillet, Phys. Rev. B {\bf 85}, 125120 (2012).

\bi{Dub09} Y. Dubi and M. Di Ventra, Nano Letters {\bf 9}, 97 (2009).

\bibitem{Dubi09b}
Y.~Dubi and M.~Di~Ventra, Phys. Rev. E {\bf 79}, 042101 (2009).

\bibitem{Dubi09a}
Y.~Dubi and M.\ Di Ventra, Phys. Rev. B {\bf 79}, 115415 (2009).

\bi{Cas10} A. Caso, L. Arrachea, and G.S. Lozano, Eur. Phys. J. B {\bf 85}, 266 (2012).

\bi{But88} M. B\"uttiker, Phys. Rev. B {\bf 33}, 3020 (1986).

\bi{Gra97} T. Gramespacher and M. B\"uttiker, Phys. Rev. B {\bf 56}, 13026 (1997).

\bi{Jac10} Ph. Jacquod, Nanotechnology {\bf 21}, 274006 (2010).

\bi{Imry86} Y. Imry, in {\it Directions in Condensed Matter Physics}, G. Grinstein and G. Mazenko eds.,
World Scientific (Singapore, 1986).
 
\bi{But90} P. N. Butcher, 
J. Phys.: Condens. Matter {\bf 2}, 4869 (1990).

\bi{San11} D. S\'anchez and L. Serra,
Phys. Rev. B {\bf 84}, 201307 (2011).

\bi{JN} J.B. Johnson, Phys. Rev. {\bf 32}, 91 (1928);
H. Nyquist,  Phys. Rev. {\bf 32}, 110 (1928).

\bi{Bla00} Ya.M. Blanter and M. Buttiker, Phys. Rep. {\bf 336}, 1 (2000).

\bi{Bee97} C.W.J. Beenakker, Rev. Mod. Phys. {\bf 69}, 731 (1997).

\bi{Ber13} J.P. Bergfield, M.A. Ratner, C.A. Stafford, and M. Di Ventra, arXiv:1305.6602.

\end{thebibliography}
\end{document}